\documentstyle[12pt]{article}
\newcommand{\be}{\begin{equation}}
\newcommand{\ee}{\end{equation}}

\def\psnormal{\textwidth=16.5cm\textheight=21.9cm
          \oddsidemargin=0.25cm\evensidemargin=0.25cm
          \topmargin=0cm\parindent=1cm}

\input epsf
\psnormal

\def\G{\Gamma}
\def\Mp{M_{P}}
\def\lh{\lambda_{H}}

\def\S{{\cal S}}
\def\G{{\cal G}}
\def\Lh{{\cal L}_{H}}
\def\Ls{{\cal L}_{MSSM}}
\def\Lo{{\cal L}_{obs}}
\def\Loh{{\cal L}_{OH}}
\def\dgs{\tilde{\delta}^{GS}}

\begin{document}

\pagestyle{empty}

\hspace{3cm}

{\vbox{\baselineskip=12pt{\rightline{\small{ OUTP--94--16P}}}}}
\vspace{2cm}
\begin{center}
{\bf SOFT SUPERSYMMETRY BREAKING FROM GAUGINO CONDENSATION}
\vspace{2cm}

B. de CARLOS\footnote{Work supported by a Spanish M.E.C. grant.} \ and \
M. MORETTI\footnote{European Community Fellow,  CHRX-CT93-0132
contract, Human Capital and Mobility program.}

\vspace{0.7cm}

 {\it Department of Theoretical Physics, University of Oxford, \\
1 Keble Rd, Oxford OX1 3NP}
\vspace{0.5cm}

\end{center}

\centerline{\bf Abstract}
\vspace{0.7cm}

{\vbox{\baselineskip=15pt
\noindent

We study the structure of soft breaking terms in the context of a gaugino
condensation scenario. Assuming that the Supergravity Lagrangian is the
correct quantum field theory limit, at some momentum scale $\mu_{UV}$, of
a more fundamental one, we demonstrate that the correct result is obtained
simply by substituting, in the tree level Supergravity Lagrangian,
$\lambda \lambda$ (the gaugino condensate) by its vacuum expectation value
$\Lambda^3$. In string inspired scenarios this implies,
in particular, that the scalar masses are vanishing at the string
tree-level and receive a contribution, at the one loop level, which is
proportional to the Green Schwarz coefficient $\delta_{GS}$. Our results
do not agree with the ones obtained in the effective Lagrangian approach.
We study in detail the origin of this discrepancy, and we argue that the
use of the supertrace anomaly to determine the effective theory for the
condensate does not fix its gravitational interactions, leaving the soft
breaking terms and the vacua of the theory unspecified.
 }}

\vspace{0.7cm}
\begin{flushleft}
{\vbox{\baselineskip=12pt {\small{OUTP--94--16P}} \\
{\small{July 1994}}}}
\end{flushleft}

\newpage
\baselineskip=16pt
\pagestyle{plain}
\pagenumbering{arabic}
\section{Introduction}
Many extensions of the Standard Model embody supersymmetry (SUSY) as a
fundamental ingredient. Since the low energy world is manifestly
non-supersymmetric, these models are specified only when the source of
SUSY breaking is defined.

The main motivation for low energy SUSY is linked to the absence of
quadratic divergencies in the perturbative expansion, which is essential
in providing a solution, at least at the technical level, to the
hierarchy problem \cite{thoof81}. If one wants to keep this important
property, the only allowed explicitly non-supersymmetric terms in the
Lagrangian are the so called soft breaking terms (SBT) \cite{girar82},
whose mass scale has to be the electroweak scale.

When SUSY is promoted to be a local symmetry of nature one obtains
Supergravity theories (SUGRA) \cite{barbi82,cremm83,nille84}. It is in this
framework in which we hope to understand the origin of SBT and their
typical scale. Furthermore, if some dynamical condensation occurs, SUSY
is broken and the non renormalizable terms of the Lagrangian origin the
wanted SBT with a typical mass scale $m_{SB}=\Lambda^3/\Mp^2$, $\Lambda$
being the scale of the condensate and $\Mp$ the Plank mass. If
$\Lambda \sim 10^{-5} \Mp$ the required hierarchy is generated. Since in
the SUGRA Lagrangian the only fundamental mass scale is $\Mp$, one expects
that $\Lambda$ is dynamically generated as the confinement scale of some
non abelian sector of the theory \cite{barbi82,nille82} which is usually
assumed to be decoupled, with the exception of non renormalizable interaction
terms, from the observable sector (the Standard Model one). A SUGRA
Lagrangian is specified when three functions of the fields are given, namely
the real--analytic K\"ahler potential $K$ and two holomorphic functions
$P$ and $f$, the superpotential and the gauge kinetic function respectively.
Since the theory is non-renormalizable, another essential ingredient is the
ultraviolet cutoff.

In the context of a fundamental theory, it is possible to try to calculate
the SBT, allowing for a non trivial test of it. Superstrings are the most
promising candidates for such a theory \cite{green86}; in this framework
there is only one fundamental parameter: the Planck scale
($M_{P} \sim 10^{18}$ GeV), and any other quantity is determined dynamically,
that is, as the vev of a field. Furthermore the theory is finite, so it has
to be anomaly free under any symmetry that it possesses. Such constraints
allow in principle for an exceptional predictive power; so, under the
assumption that SUGRA is indeed the correct field theory limit of
superstrings, one can try to make use of the available information to limit
the possible form of the SBT.

Given a particular compactification scheme, in principle, the $f$, $K$, and
$P$ functional forms can be computed. In practice this has only been done
for orbifold \cite{dixon85,casas89} and large radius Calabi--Yau
compactifications \cite{cande85}, for which we have expressions for $f$ and
$K$ at the one loop level; concerning $P$, we must require that, at least,
it contains the non--gauge standard model interactions.

Despite of all the progress done, there are still lots of problems to be
solved in order to give string theories a complete predictive power; one of
the most important is the complete understanding of the SUSY breaking process.
The most promising option is that of non-perturbative effects as the source
of SUSY breaking \cite{banks88}, and, among these, gaugino condensation is
the most appealing one \cite{nille82,ferra83,deren85,dine85}. In SUGRA, as
already discussed, non-renormalizable contributions usually appear in the
expression of the auxiliary fields $F^j$, and these terms can induce SUSY
breaking in presence of a gaugino condensate (a non trivial vev for $F^{j}$
indicates that SUSY is broken) \cite{deren85}. Requiring that the effective
potential for the gaugino condensate correctly reproduces the quantum
behaviour under both anomalous and non anomalous R symmetries, and that it
embodies modular invariance, an effective Lagrangian depending on the
condensate, the dilaton and the moduli fields has been found
\cite{ferra90,binet89} (from now on, we shall refer to this as effective
Lagrangian approach). The usual assumption which is made is that this non
pertubative contribution can be reproduced by a proper term in the effective
superpotential. Under this, all non perturbative effects, including other
possible unknown contributions, can be accounted for assuming that the
superpotential develops a non-zero vev triggering SUSY breaking
\cite{ferra90,binet89,font90}.

Having specified the mechanism for SUSY breaking together with $f$, $K$ and
$P$, it is possible to examine the structure of the resulting SBT, trying
to outline the peculiarities of the low energy spectrum. This has been done
first in \cite{cveti91}, and, more recently, in \cite{kaplu93,brign93,decar93}
the most remarkable feature being that a correlation between the gaugino
masses and squark and lepton masses has been found.

In an attempt to study the dynamics of the condensation mechanism, it has
been suggested in \cite{macor94} that the strong binding effects in the hidden
sector be parametrised by a four Fermi interaction along the lines of NJL
model. Two major differences are found with respect to the alternative
approach: a non trivial vacua even with one condensate only, and a different
structure for the SBT. Studying gaugino condensation from a slightly different
perspective from the usual one, we aim to investigate the origin of such
discrepancies.

Under the assumptions that the SUGRA Lagrangian is the correct quantum field
theory limit, at some scale, of a more fundamental one, and that gaugino
condensation is the only source of SUSY breaking, in section 2 we derive the
expressions for the SBT as functions of the tree level parameters $f$, $K$
and $P$, where by tree level we mean those which define the effective quantum
field theory and are obtained after having integrated out all the heavy
frequencies. In section 3 we apply our results to the specific case of string
inspired SUGRA. Since the results do not agree with those obtained
\cite{cveti91,kaplu93,brign93,decar93} using the effective Lagrangian approach,
in sections 4 and 5 we discuss the origin of the discrepancies and we point out
that, in our opinion, the arguments relying on anomaly cancellation do not
constrain the gravitational interactions of the condensate. This implies that
all the terms of order $\Lambda^{6}/M_{P}^{2}$ are completely arbitrary to the
extent to which the anomaly is concerned. Different guesses for these terms
lead to different results for the SBT. Finally we briefly discuss the issue of
the cosmological constant, and we come to our conclusions.

\section{Soft breaking terms}

In the forthcoming  sections we will closely follow ref.~\cite{cremm83},
which we will denote from now on as [C], and we will quote as (Cn) the (n)
equation of paper [C].

Our aim is to examine gaugino condensation effects in the framework of SUGRA
theories, relying on the minimum possible set of assumptions. Namely we
restrict ourselves to three basic assumptions:

$i)$ some "fundamental" theory of nature admits an $N=1$ SUGRA effective
quantum field theory limit at an effective ultraviolet cutoff scale $\mu_{UV}$.

$ii)$ it contains a sector (which in the following we will refer as hidden)
which can be described as an infrared confining SUSY gauge theory.

$iii)$ gaugino condensation occurs, namely a non zero vev for $\lh\lh$
develops, where by $\lh$ we denote the fermionic partners of the ordinary
gauge bosons.

It is clear that this is indeed the minimal possible set of assumptions under
which the problem we want to study is not meaningless. We will demonstrate
that this is sufficient to determine completely the contribution of gaugino
condensation effects to the SBT as a function of $f$, $K$, $P$ and the
condensate scale only. In summary, our starting point is a SUGRA theory with
a hidden, non abelian,  purely gauge sector (the extension to a hidden sector
containining matter fields is straightforward and not essential to the point
we want to make). The Lagrangian we are interested in, however, is not this
but the one we obtain after we have integrated out the hidden sector modes.
The resulting low energy theory, which in the following we will denote as
$\Ls$, will be the SUGRA (C4.17-4.20) one which is obtained setting to zero
the hidden fields ($\Lo$), plus, possibly, explicit SUSY breaking soft terms
(${\cal L}_{soft}$): $\Ls = \Lo + {\cal L}_{soft}$. The set of soft breaking
terms is given by:
\begin{eqnarray}
Trilinear \; piece & : & A_{ijk} h_{ijk} \varphi_{i} \varphi_{j} \varphi_{k} +
{\rm h.c.} \nonumber \\
Gaugino \; mass \; term & : & M_{a} \lambda_{a} \lambda_{a} \label{soft0} \\
Scalar \; mass \; term & : & m_{i}^{2} |\varphi_{i}|^{2} \nonumber
\end{eqnarray}
where the trilinear piece is derived only if a superpotential of the form:
$W_{Y} = h_{ijk} \varphi_{i} \varphi_{j} \varphi_{k}$ is present, containing
the Yukawa interactions between the observable matter fields ($\varphi_i$)
(here $h_{ijk}$ are the Yukawa couplings). We have also another soft term, a
bilinear coupling between the two Higgs fields. In general it will have the
form: $B \mu H_{1}H_{2} + {\rm h.c.}$, with $\mu$ a mass dimensional quantity
which will depend on the origin of this term. We will discuss later on the
different possibilities proposed up to now.
\begin{figure}[b]
\epsfysize=1.3in
\epsffile[-60 -30 900 60]{diag1.ps}
\begin{center}
\parbox{5.5in}{
\caption[]
{\small
 The squark propagator. $H_i$ denotes that only internal hidden fields, in a
generic instanton background, are accounted for. The set of graphs denoted by
$B$ do not contribute to wave function renormalization.
}}
\end{center}
\end{figure}

In the following, for notational convenience, we will assume our "Standard
Model" to consist of only one matter chiral superfield $\Phi$ and one gauge
sector. The scalar component of the superpotential is $P=h\varphi^3$ (where
$\varphi$ is the scalar component of $\Phi$), and the fermionic partners of
the gluons are collectively denoted by $\lambda_g$. The extension of this
discussion to a more realistic Standard Model is straightforward.

Now that we have defined our notation, we are ready to discuss the structure
of the SBT. As a first step we need to integrate out the hidden sector modes.
We find more clear and enlightening to work in the explicit component notation
[C], in which the effective Lagrangian $\Ls$ is defined by:
\be
e^{\int \Ls} = \left< e^{\int (\Lo + \Loh + \Lh)} \right>_{H} \; \; ,
\label{exp}
\ee
where $\Lh$ is the SUGRA Lagrangian (C4.17-4.20) obtained setting to zero the
observable fields, $\Loh$ denotes the whole set of gravitational interactions
among hidden and observable fields, and $<>_{H}$ denotes the functional
integration over the hidden degrees of freedom.
The relevant piece for the study of the soft breaking terms is:
\begin{eqnarray}
e^{\int \Ls} & = & e^{\int \Lo} \left< \int \left[ \frac{\bar{\lh}
\bar{\lh}}{\Mp^2} (\tau \varphi^{3} + \G \lambda_{g}  \lambda_{g})+ {\rm h.c.}
+ \S \frac{|\lh \lh|^{2}}{\Mp^{4}} |\varphi|^{2} \right]  e^{\int \Lh}
\right>_{H} \nonumber \\
& + & O \left< \left( \int \frac{|\lh \lh|^{2}}{\Mp^{6}} |\varphi|^{4} +
\frac{\bar{\lh} \bar{\lh}}{\Mp^4}|\varphi|^{2} \varphi^{3} + {\rm h.c.} +
\ldots \right) e^{\int \Lh} \right>_{H} \; \; ,
\label{exp2}
\end{eqnarray}
where $\tau$, ${\cal G}$ and ${\cal S}$ are the coefficients of the
corresponding interactions in the Lagrangian and will be given later. Notice
that the structure of $\Loh$ is, of course, richer than the one given in
(\ref{exp2}); however, by direct inspection of the SUGRA Lagrangian it turns
out that, at the relevant order of expansion ($e^{\Loh} \simeq 1 +  \Loh$),
any other possible "soft" contribution would lead to a disastrous breaking of
Lorenz--Poincar\'e invariance.
\begin{figure}[t]
\epsfysize=1.3in
\epsffile[-30 -20 2000 100]{diag2.ps}
\begin{center}
\parbox{5.5in}{
\caption[]{\small
The contribution of hidden fields to the scalar's wave function
renormalization.
Dashed lines denote scalars and the continuous one a generic hidden field.
}}
\end{center}
\end{figure}

We wish to point out that the procedure leading to (\ref{exp2}) amounts to
sum up all the contributions coming from loops of hidden particles only.
For example, the scalar mass in (\ref{exp2}) is equivalent to the "calculation"
of the scalar's propagator allowing only hidden particles in the loops. As
shown in figs.~1, 2 and 3 this is generically a sum of unknown functions and
of unknown dimensionful numbers. This inspection turns out to be useful only
because of the peculiarly simple structure of $\Loh$, indeed no coupling
provides any contribution in fig.~2 and only the coupling
${\cal S} |\varphi|^2 |\lh\lh|^2$ contributes in fig.~3. The whole non
perturbative contribution, as shown in fig.~4, can therefore be parametrized
by a single unknown mass parameter $\Lambda^3/\Mp^2$. To be more precise,
the effect of the propagation of hidden particles is also to generate higher
order (possibly non local) operators $ \sim \varphi^n$. These are, however,
non renormalizable, and therefore we are forced to neglect them to obtain a
predictive theory, hoping that the "final" theory will provide with a
justification for this assumption.
\begin{figure}[b]
\epsfysize=1.3in
\epsffile[-30 -20 2000 100]{diag3.ps}
\begin{center}
\parbox{5.5in}{
\caption[]{\small
Same as in fig.~2, but graphs contributing to the scalar's mass
renormalization only.
}}
\end{center}
\end{figure}

Assuming factorization, we have:
\be
<|\lh \lh|^{2}>_{H} = <\lh\lh>_{H} <\bar{\lh} \bar{\lh}>_{H}
\label{fact}
\ee
This has been demostrated to hold for supersymmetric theories, even when SUSY
is spontaneously broken \cite{shifm85}, if $\lh\lh$ is a gauge invariant
quantity. In the case we are discussing this is not true, due to gravitational
interactions; however, we expect that any deviation from (\ref{fact}) is
suppressed at least by loop factors of order $1/(16\pi^2)$.
\begin{figure}[b]
\epsfysize=1.3in
\epsffile[-130 -20 1000 100]{diag4.ps}
\begin{center}
\parbox{5.5in}{
\caption[]{\small
The unique contribution to the scalar's two point function in a
SUGRA theory.
}}
\end{center}
\end{figure}

We obtain that the soft breaking terms of $\Ls$ are given by:
\be
A  =  \tau \frac{\Lambda^{3}}{\Mp^{2}} \; \; ; \; \;
M  =  \G \frac{\Lambda^{3}}{\Mp^{2}} \; \; ; \; \;
m_{\varphi}^{2}  =  \S \frac{\Lambda^{6}}{\Mp^{4}} \; \; ,
\label{soft1}
\ee
where $\Lambda^{3} =  \langle \lh \lh e^{\int \Lh} \rangle_{H}$
and
\begin{eqnarray}
\tau h \varphi^3 & = & (P_{j}+K_j P) (K^{-1})_i^j \bar{f}^i\nonumber \\
{\cal G} & = & \frac {1}{32} \langle (K^{-1})_i^j f_j \bar{f}^{i} \rangle
\label {ssoft} \\
{\cal S} & = & \frac {1}{32} \left< \frac{\partial (K^{-1})_i^j}{\partial
|\varphi|^2} f_j \bar{f}^{i} \right> \; \; .\nonumber
\end{eqnarray}
One should notice that steps (\ref{exp}) and (\ref{exp2}) leading to
(\ref{soft1}) are only definitions. Nevertheless they prove to be extremely
useful in showing that, at least in the computation of the SBT, the only
information which is needed about the (non-perturbative) dynamics in the
hidden sector is the value of $\Lambda^3$. Equations (\ref{exp}) and
(\ref{exp2}) are a very simple proof of the results which are obtained
performing the "naive" substitution \cite{ferra83} $\lh\lh \rightarrow
\Lambda^3$ in the SUGRA Lagrangian.

Let us make our point explicit also in the effective potential language.
We go back to the component formulation [C], and using purely formal arguments,
we want to study the effective potential of our theory. In particular, we
need to work out a MSSM Lagrangian which contains explicitly the dependence
on the $U$ composite field ($\tilde{\cal L}_{MSSM}$). In order to do so, we
add an extra Gaussian integral to the path integral; we define
\be
\int {\cal D}U e^{- \int \tilde{\cal L}_{MSSM}} = \frac{1}{N} e^{-\int \Lo}
\int {\cal D} \varphi_{H} {\cal D}U e^{-\int (\Lh + \Loh^{'}+|U-
\frac{\lh \lh}{\Mp}|^{2})} \; \; ,
\label{pi1}
\ee
where $\varphi_{H}$ denotes a generic hidden field, $N$ is the normalization
factor which is introduced to compensate for the Gaussian integral, and
$\Loh^{'}$ contains all the relevant mixed terms between the hidden and the
observable sectors:
\be
\Loh^{'}=\S|\varphi|^2 |U|^2 + ( \tau \varphi^3 +\G \lambda_g
\lambda_g ) \frac{\bar{\lh}\bar{\lh}}{\Mp^{2}} + {\rm h.c.} + \cdots
\label{lohp}
\ee
{}From (\ref {pi1}) and (\ref {lohp}) we get the classical equation of motion
for $U$:
\be
U=\frac{\lh \lh}{\Mp} \left( 1 - \S \frac{|\varphi|^2}{\Mp^2} \right) \; \;,
\label{eqm}
\ee
where we have neglected higher order terms in $ {|\varphi|^2}$. This shows
\cite{gross74} that, indeed, $\tilde{\cal L}_{MSSM}$ is the Lagrangian we are
looking for. From the expressions (\ref{pi1}), (\ref{lohp}) and (\ref{eqm}) it
is clear that, after performing the integration over the hidden fields, we end
up with
\begin{eqnarray}
\tilde{\cal L}_{MSSM} & = & \Lo + K \left( U-\tau \frac{\varphi^3}{\Mp} -\G
\frac{\lambda_g \lambda_g}{\Mp} \right) + \left(1 + \S
\frac{|\varphi|^2}{\Mp^2}
\right) |U|^2 \nonumber \\
& + & V \left( U-\tau \frac{\varphi^3}{\Mp} -\G \frac{\lambda_g\lambda_g}{\Mp}
\right) \; \; ,
\label{veff}
\end{eqnarray}
where $K(U)$ is a kinetic functional for $U$, and $V$ is the contribution to
the effective potential, $V_{eff}=-|U|^2 - V(U)$, induced by the interaction
of the other hidden fields with $U$. It is immediately seen, by inspecting
(\ref{veff}), that $\tilde{\cal L}_{MSSM}$ cannot be obtained, in the standard
SUGRA framework, by a mere shift of the superpotential by a term accounting
for gaugino condensation. From (\ref{veff}), minimizing $V_{eff}$ with respect
to $U$, we get: $\langle \bar U \rangle = - \langle {d V}/{d U}|_0 \rangle$.
The soft breaking terms are obtained expanding the effective potential
$V_{eff}$ around the minimum $U_0$. In this case,
\be
- V_{eff}  =  |U_0|^2 \left( 1+\S\frac{|\varphi|^2}{\Mp^2} \right)  +
\frac {1}{\Mp}
 \bar{U}_0  (\tau \varphi^3 +\G\lambda_g\lambda_g ) + {\rm h.c.} + \cdots
\label{soft5}
\ee
Eq.~(\ref{soft5}) is consistent with expressions (\ref{soft1}). We stress that
the above equation is obtained with the only underlying assumption that the
standard SUGRA Lagrangian is an effective field theory below the
compactification scale. Notice that, in the above expression, we automatically
obtain the properties previously referred as decoupling (\ref {fact}). This
is a consequence of the fact that in the definition which we have given of
$\tilde{\cal L}_{MSSM}$ (see eq.~(\ref{pi1})) no integration over the $U$
field is assumed and, therefore, all the effects of its propagation are
neglected.

\section{Application to a string inspired scenario}

In order to give explicit expressions for the soft breaking terms, we need
to specify the form of the functions which define our SUGRA Lagrangian in
four dimensions, namely the K\"ahler potential, $K$, the gauge kinetic
function, $f$, and the superpotential, $P$, all of them depending on the
chiral superfields ($K$ is real--analytic, while $f$ and $P$ are holomorphic
in these fields). We are interested in effective field theories derived from
a higher dimensional string theory, for which $K$, $f$ and $P$ are completely
determined given a compactification scheme, although in practice they are only
sufficiently known for orbifold compactifications \cite{dixon85,casas89} (and
for some Calabi--Yau spaces also \cite{cande85}). In this framework we have:
\begin{eqnarray}
f^{a} & = & k^{a} S + \epsilon^{a} \log(\eta^{2}(T)) \nonumber \\
K & = &
  -\log(Y) - 3\log(T+\bar{T}) + K_{1}^{i}(T, \bar{T}) |\varphi_{i}|^{2}
 \label{fK} \\
P & = & h_{ijk}(T) \varphi_{i} \varphi_{j} \varphi_{k} \nonumber
\end{eqnarray}
where $\epsilon^a =  b^{a} - k^{a} \dgs $, $Y = S+\bar{S} +
\dgs\log(T+\bar{T})$, $S$ and $T$ are the dilaton and modulus fields
respectively, $k^{a}$ are the Kac--Moody levels and $b^{a}$ the one--loop
beta function coefficients associated to the gauge groups $G^{a}$,
$n_{\varphi_{i}}$ are the modular weights associated to the formerly defined
matter fields $\varphi_{i}$, $\dgs=\delta^{GS}/4\pi^{2}$, where $\delta^{GS}$
is the usual Green--Schwarz coefficient \cite{deren92} and $\eta(T)$ is the
Dedekind function ($\eta(T) = e^{-\frac{\pi T}{12}} \prod_{n=0} (1-e^{-2 \pi
n T})$). Finally, in order to simplify the notation, we will take a generic
matter field $\varphi$ and assume for the superpotential the following
simplified expression: $P = h(T) \varphi^{3}$, where $h(T)$ is a Yukawa
coupling.

{}From now on we will work only with one gauge group $G$ in the hidden sector
and the corresponding Kac--Moody level $k=1$, being the generalization of our
results for more than one gauge group and non trivial levels completely
straightforward. Also the dependence of the K\"ahler potential $K$ in the
matter fields is known as a series expansion \cite{deren92}. In what concerns
to our calculation, it is enough to keep terms up to $|\varphi|^{2}$ and
neglect the rest. The coefficient $K_{1}$ in (\ref{fK}) is given by:
$K_{1}(T, \bar{T}) = (T+\bar{T})^{n_{\varphi}}$, where we are again
considering a single generic matter field $\varphi$ with modular weight
$n_{\varphi}$.

Now, from the expression of the SUGRA Lagrangian [C], and taking into account
the factorization property (\ref{fact}) and eq.~(\ref{soft1}), we can derive
the soft breaking terms for this particular case:
\begin{eqnarray}
M & = & \frac{1}{32 {\rm Re}f} \left[ Y^{2} + \frac{Y}{(3Y+\dgs)} \left| \dgs -
2 \epsilon (T+\bar{T}) \frac{\eta^{'}(T)}{\eta(T)} \right|^{2} \right]
\frac{\Lambda^{3}}{\Mp^{2}} \nonumber \\
A & = & \frac{-Y^{1/2}}{4(T+\bar{T})^{3(n_{\varphi}+1)/2}} \left[ 1 - \frac{1}
{3Y+\dgs} (3(n_{\varphi}+1)  - (T+\bar{T})\frac{ h_{T}(T)}{h(T)})
 \right. \nonumber \\
  & \times & \left. \left( \dgs - 2(T+\bar{T})\epsilon
\frac{\eta^{'}(T)}{\eta(T)} \right) \right] \frac{\Lambda^{3}}{\Mp^{2}}
\label{soft3} \\
m_{\varphi}^{2} & = & \frac{-n_{\varphi}}{32} \frac{Y^{2}}{(3Y+\dgs)^{2}}
\left| \dgs - 2 (T+\bar{T}) \epsilon \frac{\eta^{'}(T)}{\eta(T)} \right|^{2}
\frac{\Lambda^{6}}{\Mp^{4}} \nonumber
\end{eqnarray}
where $h_{T} = \partial h/\partial T$ and $\eta^{'} = \partial \eta/\partial
T$. Here we have included the normalization for matter fields and gauginos
due to non canonical kinetic terms.

The most remarkable comment which is in order is that scalar masses vanish in
the limit in which no one--loop corrections to $K$ and $f$ are taken into
account (that is, when $\dgs=0$). Notice that, to have scalar masses of the
same order of magnitude of the gaugino ones, a large $\dgs$ coefficient and/or
a large modular weight for the field $\varphi$ is needed.

We now turn out to discuss the $\mu$ and $B$ terms; it is known that a
coupling of the form $\mu H_{1} H_{2}$ has to be present in the matter
superpotential in order to generate the correct ${\rm SU}(2)_{L} \times
{\rm U}(1)$ breaking; furthermore, $\mu$ should be of the order of the SUSY
breaking scale, and the soft term which generates in the Lagrangian is of the
form: $B \mu H_{1} H_{2} + {\rm h.c.}$ \cite{kim84}. Its calculation is
analogous to that of the trilinear term:
\begin{eqnarray}
B \mu & = & \frac{-Y^{1/2}}{4(T+\bar{T})^{(3+n_{1}+n_{2})/2}} \left[\mu -
\frac{1}{3Y+\dgs} ((3+n_{1}+n_{2})\mu- (T+\bar{T}) \mu_{T}) \right. \nonumber
\\
& \times & \left. \left( \dgs- 2 (T+\bar{T}) \epsilon
\frac{\eta^{'}(T)}{\eta(T)} \right) \right] \frac{\Lambda^{3}}{\Mp^{2}}
\label{B}
\end{eqnarray}
where $\mu_{T} = \partial \mu/\partial T$.

On the other hand, it has been suggested an alternative origin for this
coupling \cite{casas93}, namely the presence in the K\"ahler potential of
terms like:
\be
K^{'} = K + \nu(S,T) H_{1} H_{2} + {\rm h.c.} \; \; .
\ee
In this case the soft term is induced in the same way as the scalar masses do,
so that we end up with a $B$ term of the form:
\begin{eqnarray}
B  & = & \frac{-1}{32 \mu } \frac{Y^{2}}{(3Y+\dgs)^{2}} (T+\bar{T})^{\frac
{2-(n_{1}+n_{2})}{2}} \left[ \nu_{T}^{\bar{T}}(T+\bar{T}) -
\nu_{T}(n_{1}+n_{2}) \right] \nonumber \\
& \times & \left| \dgs - 2 (T+\bar{T}) \epsilon \frac{\eta^{'}(T)}{\eta(T)}
\right|^{2} \frac{\Lambda^{6}}{\Mp^{4}} \; \; ,
\label{Bp}
\end {eqnarray}
where $\nu_{T}(^{\bar{T}}) = \partial{\nu}/\partial{T}(\bar{T})$ (we have only
considered a possible $T$-dependence of $\nu$).

\section{Gravitational interactions of the condensate}

Since the expressions we have got for the soft breaking terms disagree with
the ones computed  using the effective Lagrangian approach
\cite{cveti91,kaplu93,decar93}, it is useful to study in detail the results
obtained in this latter framework, and try to understand the origin of the
discrepancy. We shall closely follow the approach of ref.~\cite{binet89},
which in the following we will always quote as [B], and we will refer as (Bn)
the equation (n) of [B].

In the notation of [B] the SUGRA Lagrangian (B3.7), (B3.10), (B3.11) is given
by ${\cal L} = {\cal L}_{0} + {\cal L}_{YM} + {\cal L}_{pot}$ where:
\begin{eqnarray}
{\cal L}_{0} & = & -3 \int d^{2} \Theta {\cal E} {\cal R} + {\rm h.c.}
\nonumber \\
{\cal L}_{YM} & = & \frac{1}{4g^{2}} \int d^{2} \Theta {\cal E} f(\Phi) U +
{\rm h.c.}  \label{lag} \\
{\cal L}_{pot} & = & \int d^{2} \Theta {\cal E} e^{K/2} P(\Phi) + {\rm h.c.}
\nonumber
\end{eqnarray}
where $U = \frac{1}{4} W^{\alpha} W_{\alpha}$, and $W_{\alpha}$ is the chiral
gauge supermultiplet (for the notation, see [B]).

Using these expressions we obtain for the component Lagrangian (${\cal
L}_{s}^{c}$), which contains only scalar interactions, the following equation:
\be
{\cal L}_{s}^{c} = F_{j} ({\cal G}^{-1})_{i}^{j} \bar{F}^{i} - 3 e^{{\cal G}}
\; ,
\label{int}
\ee
where $ {\cal G} = K + \log |P|^{2}$ and
\be
F_{j} = - e^{K/2} [ K_{j} P + P_{j} ]  + \frac{1}{4} f_{j} U \; \;.
\label{Fu}
\ee
This generates three different types of terms:
\begin{eqnarray}
& i) & e^{K} [(K_{j} P + P_{j}) (K^{-1})_{i}^{j} (K^{i} \bar{P} + \bar{P}^{i})
- 3|P|^{2}] \nonumber \\
& ii) & \frac{-1}{4} e^{K/2} (K_{j} P + P_{j}) (K^{-1})_{i}^{j} \bar{f}^{i}
\bar{U} + {\rm h.c.}  \label{terms} \\
& iii) & \frac{1}{32} f_{j} (K^{-1})_{i}^{j} \bar{f}^{i} U \bar{U} \; \; .
\nonumber
\end{eqnarray}
In our particular case, we are assuming gaugino condensation as the only
source of SUSY breaking, so we take $<P>=0$, and a superstring inspired
scenario, in which both $f$ and $K$ are going to depend on $S$ and $T$. So
we end up with $iii)$ in eq.~(\ref{terms}) as the only surviving interaction,
with $i,j=S,T$.

According to [B], we can rewrite (\ref{lag}) by making the following shift:
\begin{eqnarray}
{\cal L}_{YM}^{'} & = & 0  \nonumber \\
{\cal L}_{pot}^{'} & = & \int d^{2} \Theta {\cal E} e^{K/2} P^{'}(\Phi) +
{\rm h.c.}  \label{KPf2} \\
P^{'} & = & P + \frac{1}{4} e^{-K/2} f U   \; \; . \nonumber
\end{eqnarray}
As discussed in [B], this is going to introduce a non holomorphic piece in
the superpotential $P^{'}$. To recast it in an holomorphic way, in [B] a new
superfield $H$ is introduced with the proper K\"ahler transformation: $U =
e^{K/2} \rho(S) H^{3}$ (see eq.~(B4.16)). In terms of this $H$ field the
superpotential $P^{'}$ reads:
\be
P^{'} = P + \frac{1}{4} f \rho(S) H^{3}
\label{Pp}
\ee
(see eq.~(B4.21), with $\rho$ instead of $f$ and $f$ instead of $S$).

Let us note that all these formal steps are perfectly consistent with the
required K\"ahler invariance of the theory. However, the previous field
redefinition does not correspond only to a simple reparametrization of the
field $U$ in terms of $H$; it is clear from (B4.15-4.17) that $H$ and $U$
have different conformal weights, as explicitly noticed in [B] and, therefore,
they are expected to have different gravitational interactions.

Now that the effects of gaugino condensation have been included in the
superpotential, we go back to the component notation and look for the scalar
interactions in this new language. The F term is now:
\be
F^{'}_{j}  = - e^{K/2} \left[ K_{j} P + P_{j} - \frac{1}{4} (K_{j} f \rho(S) +
f_{j} \rho(S) + f \rho_{j}(S) ) H^{3} \right] \; \; .
\label{Fh}
\ee
The crucial difference between (\ref{Fu}) and (\ref{Fh}) is given by the
appearance of the $e^{K/2}$ factor in front of the $H$ field in this latter
equation, which is a consequence of considering this field as a fundamental
dynamical variable\footnote{In fact, we could formally rewrite (\ref{Fh}) with
$P^{'}$ given in terms of the $U$ field as defined in eq.~(\ref{KPf2}): the
$e^{K/2}$ factor would disappear and we would recover expression (\ref{Fu}).
Notice that this is not the case for the theory in terms of the $H$ field:
substituting back $U$ by $H$ in (\ref{Fu}) does not give the same result as
(\ref{Fh}).}.

Given the former expression for $F^{'}_{j}$, the part of the Lagrangian which
contains scalar interactions, that is (\ref{int}), will consist now of a term
analogous to $i)$ in eq.~(\ref{terms}), with $P^{'}$ instead of $P$:
\begin{eqnarray}
{\cal L}_{s}^{c} & = & e^{K} \left[ \frac{1}{32} (K_{j}\rho(S) f+\rho_{j}(S) f
+ \rho(S) f_{j}) (K^{-1})_{i}^{j} (K^{i} \bar{\rho}(\bar{S}) \bar{f} +
\bar{\rho}^{i}(\bar{S}) \bar{f} + \bar{\rho}(\bar{S}) \bar{f}^{i}) |H^{3}|^{2}
\right. \nonumber \\
& - & \left. 3 |\rho(S)fH^{3}|^{2} \right] \; \; , \; \; k, l = S, T \; \; ,
\label{Ls2}
\end{eqnarray}
where we again stress the presence of the global $e^{K}$ factor multiplying
the condensate.

In addition we notice a more subtle difference arising between the two
formulations. The fact that the SUGRA Lagrangian is not positive definite
is due to a factor $-3|P|^{2}$, which appears in the scalar potential
(eq.~(\ref{terms}) $i)$) after substituting the auxiliary field $\omega$ of
the Weyl compensator superfield, introduced in order to give the right
conformal weight to the fields, as can be checked by looking at (C4.3-4.5)
(with $u$ instead of $\omega$). Given that the $U$ and $H^{3}$ fields have
different conformal weights, the dependence of $\omega$ on $H^{3}$ and $U$ is
completely different; in particular, again from a direct inspection of the
component Lagrangian, one can check that using the $H$ language a factor
$-3 |\rho(S) f H^{3}|^{2}$ is present, which has no corresponding piece in
the $U$ language.

While it is not clear to us how all these differences affect the problem of
minimizing the potential, they completely change the structure of the SBT. We
conclude that the usual assumption that all the non--perturbative effects can
be taken into account shifting the tree--level potential by an amount which
depends only on the composite field leads to an incorrect conclusion.

Notice that, since all the steps done in [B] to go from the $U$ to the $H$
formulation always lead to a Lagrangian with the correct K\"ahler
transformation, it should be true also for the inverted one. Given that the
two formulations lead to different conclusions, we guess that the arguments
relying on anomaly cancellation to inspect the possible behaviour of the
effective potential suffer from some ambiguites. Indeed it seems that
physically inequivalent potentials have the same properties under both the
anomalous transformation (B4.3), (B4.4) and the nonanomalous one (B4.7-B4.9).
We notice that, motivated by a different purpose, Bin\'etruy and Gaillard
reached essentially the same conclusion in \cite{binet91} and a somewhat
related discussion was developped in \cite{nille90}. Finally we think that a
more extreme option could be possible: namely it might be that higher order
effects cannot be incorporated in the standard SUGRA Lagrangian without
including higher order derivatives.

\section{Cutoff dependence of the effective theory}

The effective theory we are looking for depends critically on its
ultraviolet cutoff which, as any other quantity, is generically a
field dependent one. In string theory the cutoff is twofold: $i)$
at the string scale $M_S$ the particles of the theory begin to feel
 their non-pointlike structure; $ii)$ at the compactification
 scale $M_C$, the low energy degrees of freedom feel
 the interactions with the infinite tower of heavy Kaluza
 Klein modes, generated by
the compactification of the extra space dimensions.

Clearly the effects induced by $M_C$ can lead to some field
dependence. Indeed,
it may very well be that $M_C$ is a dynamical variable, and that
the compactification to four space time dimensions arises because
it is
energetically favoured \cite{taylo88}.
Moreover, all the interactions of the light degrees of freedom
with the heavy modes induce a coupling dependent effective
cutoff $\mu_{UV}$, which consequently implies its field dependence.
In string theory the field dependence of the cutoff is
 somewhat understood [B], however it is not clear to us whether
 it is controlled with the level of accuracy which is needed for
 the purpose here discussed.

We believe that this possible ambiguity is not reflected in
 the study of the soft breaking terms. Considering it from the
 low energy theory point of view, the
effective cutoff has to be a pure number, which means that it
 has to be the vev of some background non-propagating
field like the $S$ and $T$ fields. What might happen is that
the same interactions which give rise to the cutoff actually
 modify the
couplings of the low energy effective theory. To the extent
that this is a SUGRA theory, the only possible
 modification affecting the low energy modes is through
 a modification of $f$, $K$ and $P$. Notice that, in fact, this
 is what happens in the cases where high energy modes' effects
 are known, namely these
induce a modification both of $f$ and $K$, leaving $P$
invariant according to the non--renormalization theorem. The
$S--T$ mixing term, which is the origin of the scalar masses
we found in (\ref{soft3}), is indeed a manifestation of
 this mechanism: the same interactions which act as an
effective cutoff for the theory originate an effective
 $|\varphi|^2 |\lh\lh|^2$ coupling which is not present at
the string tree level.
What it is not completely clear to us is if the threshold
 effects computed up to now \cite{deren92,kaplu88} are
general enough. In quantum field theory it is well known
 that Yukawa type interactions do not affect (at one loop)
 the renormalization
 of gauge couplings, however they induce
a non trivial wave function renormalization. If this case has some
parallel in superstrings,
the K\"ahler potential might be affected inducing an
 extra dependence
on $|\varphi|^2 $ in $K_S^{\bar{S}}$ which would modify
 the structure of $m^2_\varphi$.

Finally, let us notice an important point related to the
 cutoff dependence
of the effective potential. We have stated that we regard
 this problem as
a serious one for any attempts of studying the minima of
the potential relying on the effective Lagrangian approach.
 In addition
to a possible arbitrary dependence of $\mu_{UV}$ on $S$ and
 $T$, which
perhaps can be somehow controlled using symmetry arguments, in
 our opinion
a more serious problem arises. Notice that all the discrepancies
among the approach we are suggesting and the
effective Lagrangian one are of order
 $\delta^2=\Lambda^2/\Mp^2$, and
that to study the minima in the $S$ and $T$ directions, using
the effective Lagrangian, these terms are essential.
Now, we believe that, in doing so, one should pay attention to
the fact that the
one loop super trace anomalies, which are the building blocks
 for the effective Lagrangian approach, are computed in the
 field theoretical
limit, namely in the limit of infinite cutoff. Consistently,
order $\delta^2$ terms should be neglected or, alternatively,
one should spell out the supertrace anomalies
 including $1/\mu_{UV}^2$ corrections, which does not seem
 easy without having a detailed understanding of how the
 effective cutoff works.

Let us explain this point more in detail.
In the effective Lagrangian approach one constructs an
 effective Lagrangian
${\cal L}_{eff}$ for the
 condensate $\phi=\beta(g)/(2g)\lambda\lambda$ which reads
 as follows \cite{venez84}
\be
{\cal L}_{eff} = \alpha^{-1}(\phi^*\phi)^{-2/3} \partial_\mu \phi^*
\partial_\mu \phi - \frac{1}{9} \alpha (\phi^*\phi)^{2/3}
\left| \log \frac{\phi}{\mu_{UV}^{3}} \right|^2 \; \; ,
\label{lageff}
\ee
which, properly rescaling the field $\phi$,
reads as
\be
{\cal L}_{eff} =  \partial_\mu \hat \phi^*
\partial_\mu \hat \phi -
\frac{\alpha^{3}}{81}
 \left| \hat\phi^2\log \frac{\sqrt{\alpha}\hat\phi}
{3\mu_{UV}} \right|^2
\label{lageff1}
\ee
with $\hat \phi = (3/\sqrt \alpha) \phi^{1/3}$, and $\alpha$
 a constant.

Having done this, one incorporates \cite{ferra90,binet89}
 ${\cal L}_{eff}$ in a SUGRA formalism, adding to
 the superpotential a term accounting
for (\ref{lageff}) and its supersymmetric counterpart, and
 adding to the K\"ahler potential a term generating the
 kinetic piece.
After expanding the SUGRA Lagrangian, one obtains
 again (\ref{lageff1})
plus additional gravitational interactions that,
with respect to (\ref{lageff1}), are suppressed by a
 factor $\delta^2$.
The leading contribution to the expression for the
 gaugino condensate
is obtained by setting
\be
\left| \hat\phi^2\log \frac{\sqrt{\alpha}\hat\phi}
{3\mu_{UV}} \right|^2 = 0 \; \; .
\label {minimum}
\ee
Now comes the problem: to study the minima in the $T$ and $S$
directions or the SBT, one has to rely on the
subleading contributions
since the leading one is vanishing. At this level we believe that
(\ref{lageff1}) is likely to be not accurate enough. Indeed,
 we think that
\be
{\cal L}'_{eff} =  \partial_\mu \hat \phi^*
\partial_\mu \hat \phi -
\frac{\alpha^{3}}{81}
 \left| \hat\phi^2\log \frac{\sqrt{\alpha}\hat\phi}{3\mu_{UV}}
 + \epsilon_1 \frac {\hat \phi^3}{\mu_{UV}} \right|^2
\label{lageff2}
\ee
is indistinguishable from ${\cal L}_{eff}$ to the accuracy
at which the divergence of the anomalous current is known.
If we derive the superpotential from (\ref{lageff2}) the
 resulting SUGRA Lagrangian contains an extra
contribution, proportional to $\epsilon_1^2$.
Terms of this order are essential when studying the
minimum equations for $S$ and $T$ and computing the SBT.
This observation, together with the discussion of the
 previous section,
leads us to conclude that the effective Lagrangian approach, as
 it stands now,
is not adequate to fix the gravitational interactions of
the gaugino
condensate. Indeed the "anomaly driven" effective Lagrangian
 is accurate
only up to terms of order $1/\Mp^2$, which are completely
 arbitrary to
the extent that the anomaly is concerned.

Having stated the problem we can now stress that
the SBT are indeed the first known information about the
gravitational interaction of the condensate.
The effective Lagrangian should reproduce the results we obtained
in (\ref{soft1}). The problem of whether these constraints
(and, possibly, others following from analogous considerations)
are sufficient to fix the ambiguities we were pointing out
is still an open question.
If this approach should work the outcome would be
extremely interesting:
one, in fact, would get a non trivial insight about the
dependence of
the theory on the physical cutoff. We plan to come back to
 this point in
a separate publication.

If this is the case, one has to go back and find the origin of
the discrepancies
between the SBT in (\ref{soft3}) and the
 ones \cite{kaplu93,brign93} obtained using the $F_{S,T}$
 dominance hypothesis. We guess that the answer has to be
that $F_\varphi$ becomes as important as  $F_{S,T}$,
and contributes to the SBT in a peculiar way which is
 not parametrizable using  $<F_{S,T}>$ only, as suggested
 in \cite{brign93}.

Up to now we have neglected the issue of the cosmological
 constant. We now briefly comment about this point. From
 eq.~(\ref {soft5})
it is clear that the natural order of magnitude for
 the cosmological
constant $V_C$ is $m_{SB}^{2} \Mp^2$, $m_{SB}$ being the scale
 of SUSY
breaking in the observable sector. There are several possibilities
which cannot be discarded without a "theory" for $V_C$ which
 is presently unavaliable:

\noindent $i)$ The cosmological constant is unrelated to the
 value of the effective potential at the minimum. In this case
 it is conceivable that,
if gaugino condensation is the only mechanism for SUSY breaking,
the discussion of the former sections is unchanged.

\noindent $ii)$ Due to some unknown effect
the effective potential of (\ref{soft5}) is vanishingly small
at the minimum. Again the discussion of the previous sections
 is unchanged.

\noindent $iii)$ The extra term needed to cancel $V_C$
comes from a non vanishing VEV ($P_O$) for the superpotential.
 In this case new contributions have to be added to the SBT.

\section{Conclusions}

In this paper we have addressed the issue of computing the
 soft breaking
terms (SBT) for a generic Supergravity Lagrangian assuming
 that gaugino
condensation is the origin of supersymmetry breaking. This
 is usually
done under the assumption that the effects of gaugino
 condensation
can be described by a proper superpotential for which an ansatz
 has been proposed \cite{ferra90,binet89}. We have tried to
 obtain our results avoiding assumptions and ansatze.

Our starting point was a SUGRA effective quantum field theory
 with ultraviolet cutoff $\mu_{UV}$ and a hidden sector
 where strong binding effects lead to  gaugino
 condensation. Working in the component formulation
 and integrating,
in a purely formal way, over the hidden degrees of freedom, we
 have firstly found expressions for the SBT, which explicitly
 break SUSY in the observable sector of the theory, in terms
 of $f$, $K$, $P$
and the vev of the gaugino condensate ($\Lambda$),
the only dynamical (non perturbative)
information about the hidden sector which is needed. We have
also applied our
procedure of integrating out the hidden sector modes
(which we stress again is purely formal, no particular
estimation of non perturbative effects is claimed) to work out
an "effective potential" for the condensate which, in spite
 of being defined
in terms of an unknown function, allows us to compute
 SBT consistent with the ones we already obtained.

Secondly, we have discussed the form of these expressions for
the case of string inspired SUGRA, i. e., assuming that we
 are dealing with an effective field theory derived from a
 higher dimensional string theory. This allowed us to use the
known expressions of $f$ and $K$ ($P$ containing just the
 Standard Model interactions) for orbifold compactifications,
to obtain gaugino and scalar masses and trilinear and
 bilinear couplings. The results show that, in the absence
 of threshold corrections to $f$ and $K$, scalar masses vanish.
These expressions are different from those obtained in the
effective Lagrangian approach and, therefore, we turned out
 to examine both formulations in order to find the source of
 the discrepancy. We point out that the usual procedure
 of incorporating the gaugino bilinear in the superpotential
forces its reparametrization in a way
that changes its gravitational interactions, which
are going to be crucial for the computation of the SBT,
possibly originating the previously mentioned differences
between the two formalisms. This makes us think that having
 a Lagrangian with the correct K\"ahler transformation and
 the right properties with respect to the anomalous
 and non--anomalous symmetries, is not enough to determine
 it unambiguously.

We discuss, as well, the role of the ultraviolet cutoff and
its possible field dependence, and conclude that, although a
 field dependent cutoff may modify the study of the
 vaccum structure of our theory in an unknown way, it can
 affect the structure of the SBT only through a modification
of $f$ and $K$, namely through threshold effects, as happens
for orbifold compactifications. An important point in
this discussion is that the supertrace anomalies are only known
up to $\Lambda/\Mp$ corrections. Higher order effects of this
 type would affect in an essential way the ansatz for the
effective SUGRA Lagrangian, opening perhaps a way of matching
 the results obtained in the two approaches.
We finally comment very briefly about the possible modification
of our results when the cosmological constant issue is taken
 into account.

\section*{Acknowledgements}

We are indebted to G.G. Ross
for suggesting us to start this investigation and for
 many enlightening discussions. We also thank P. Bin\'etruy for
 a very useful discussion.
M.M. wishes to thank the Italian INFN for financial
support in the very early stage of this work.

\end{document}